\documentclass[conference]{IEEEtran}
\IEEEoverridecommandlockouts

\usepackage{amsmath,amssymb,amsfonts}
\usepackage{algorithmic}
\usepackage{graphicx}
\usepackage{textcomp}
\usepackage{xcolor}
\usepackage[sorting=none]{biblatex}
\DeclareMathOperator{\Tr}{Tr}
\def\BibTeX{{\rm B\kern-.05em{\sc i\kern-.025em b}\kern-.08em
    T\kern-.1667em\lower.7ex\hbox{E}\kern-.125emX}}

\title{Training Quantum Boltzmann Machines with Coresets}

\author{
\IEEEauthorblockN{
Joshua Viszlai\IEEEauthorrefmark{1}, 
Teague Tomesh\IEEEauthorrefmark{2}\IEEEauthorrefmark{3},
Pranav Gokhale\IEEEauthorrefmark{3},
Eric Anschuetz\IEEEauthorrefmark{4}\IEEEauthorrefmark{3},
Frederic T. Chong\IEEEauthorrefmark{1}
}
\IEEEauthorblockA{\IEEEauthorrefmark{1}University of Chicago}
\IEEEauthorblockA{\IEEEauthorrefmark{2}Princeton University}
\IEEEauthorblockA{\IEEEauthorrefmark{3}Super.tech, a division of ColdQuanta Inc.}
\IEEEauthorblockA{\IEEEauthorrefmark{4}MIT}
}
\begin{document}

\maketitle

\begin{abstract}
Recent work has proposed and explored using coreset techniques for quantum algorithms that operate on classical data sets to accelerate the applicability of these algorithms on near-term quantum devices. We apply these ideas to Quantum Boltzmann Machines (QBM) where gradient-based steps which require Gibbs state sampling are the main computational bottleneck during training. By using a coreset in place of the full data set, we try to minimize the number of steps needed and accelerate the overall training time. In a regime where computational time on quantum computers is a precious resource, we propose this might lead to substantial practical savings. We evaluate this approach on 6x6 binary images from an augmented bars and stripes data set using a QBM with 36 visible units and 8 hidden units. Using an Inception score inspired metric, we compare QBM training times with and without using coresets.
\end{abstract}

\section{Introduction}




Recent years have seen a slowdown in the exponential improvements due to Moore's Law and Dennard scaling. This slowdown has been accompanied by a corresponding increase in attention paid to non-traditional, post-Moore's Law computer architectures including analog and neuromorphic computing. Quantum computing is another such architecture which exploits quantum mechanical properties such as entanglement and superposition to perform computations. For certain tasks, quantum computers are conjectured to outperform their classical counterparts precisely because they have access to purely quantum phenomena~\cite{feynman1982simulating}.

The magnitude of these speedups vary from exponential --- for example the factorization of large primes~\cite{shor1999polynomial}, simulating the physics of entangled systems~\cite{lloyd1996universal}, and solving linear systems of equations~\cite{harrow2009quantum} --- to quadratic improvements for unstructured search~\cite{grover1996fast}. Additionally, an area that has exploded with recent research~\cite{biamonte2017quantum, broughton2020tensorflow, rudolph2020generation} is quantum machine learning (QML). There are many different types of quantum neural networks, and while the exact nature of the quantum speedup (if it exists) for some QML algorithms is unknown, recent work suggests that applications targeting quantum data are a promising direction~\cite{kieferovaTomographyGenerativeTraining2017, huang2021quantum}.

Current quantum computers (QCs) are built from a variety of different technologies, enabling research in the early steps of realizing quantum algorithms and evaluating QCs~\cite{arute2019quantum, nam2020ground, tomesh2022supermarq}. Unfortunately there is a gap that currently exists between the capabilities of these machines and the resource requirements for many quantum algorithms. Current QCs are known as Noisy Intermediate-Scale Quantum (NISQ) devices~\cite{preskill2018quantum}. Their limited size and gate fidelities render them unable to implement the error correcting codes that are needed to implement most known quantum algorithms. Prior work has found that a co-design approach to QC system design, characterized by the breaking of abstraction layers and the sharing of information up and down the stack, can result in significantly improved performance~\cite{shi2020resource, tomesh2021quantum}.

QML applications are especially hindered by the need to load large data sets onto small quantum devices. Access to a quantum random access memory would allow a QC to coherently load quantum states representing classical data. However, it is likely that the construction of a quantum RAM is equally or more difficult than building a fault tolerant QC~\cite{arunachalam2015robustness}. Instead, prior work has investigated the use of coresets, a succinct summarization of a larger data set \cite{mirzasoleiman2020coresets}, to apply QML models to large data sets using small quantum computers~\cite{harrowSmallQuantumComputers2020, tomeshCoresetClusteringSmall2020}.

The Quantum Boltzmann Machine (QBM) is a physically motivated quantum neural network that can be used for generative or discriminative learning~\cite{aminQuantumBoltzmannMachine2018}. Prior work has studied the application of QBMs to tasks such as image generation \cite{anschuetzNearTermQuantumClassicalAssociative2019}, and they have been shown to outperform classical Boltzmann machines for certain tasks such as quantum state generation~\cite{kieferovaTomographyGenerativeTraining2017}. 

The downside of such a powerful and versatile QML model is the overhead costs associated with the training process. The gradient updates that are needed to tune the model's parameters require samples taken from a thermal (Gibbs) state
\begin{equation}
    \rho(\beta) = \frac{e^{-\beta H}}{\text{Tr}(e^{-\beta H})}
\end{equation}
where $\beta=1/T$ is the inverse temperature and $H$ is the system Hamiltonian describing the QBM. This is an NP-Hard problem \cite{kieferovaTomographyGenerativeTraining2017, anschuetz2019realizing} and to successfully train a QBM many such states will need to be prepared and sampled from.

To circumvent the difficulties of training a QBM many different techniques have been suggested. For example, rather than training on the exact loss function it can be simpler and more efficient to train on an upper bound of the loss or restrict the connectivity of the QBM model~\cite{aminQuantumBoltzmannMachine2018}.
Similarly, a variety of different quantum algorithms have been proposed for the specific task of thermal state preparation including quantum walks \cite{temme2011quantum}, quenches~\cite{anschuetz2019realizing}, semi-definite programming~\cite{brandao2017quantum}, and hybrid variational algorithms~\cite{verdonQuantumAlgorithmTrain2019}.

This work presents a complementary method for potentially reducing the training overhead of QBMs. We propose building coresets of the training data set to reduce the overall number of Gibbs state preparations needed during the training process. We run initial numerical simulations with QBMs needing 44 qubits and present an augmented bars and stripes data set and corresponding Inception score inspired metric to compare training with and without coresets.

Our contributions include
\begin{enumerate}
    \item A numerical study motivating the potential of coresets for QBM training.
    \item An augmented bars and stripes data set to evaluate generative machine learning models of moderate dimensionality.
    \item Numerical experiments verifying successful training of QBMs on this data set.
    \item Initial numerical experiments exploring the effectiveness of coresets for QBMs of moderate size.
\end{enumerate}


\section{Background}
\subsection{Coresets}
A coreset of dataset $\boldsymbol{X} = \{\boldsymbol{x_1}, ..., \boldsymbol{x_n}\}$ is a subset $\boldsymbol{X}'$ with weights $w$ such that $(\boldsymbol{X}',w)$ can be used as a proxy for $\boldsymbol{X}$ when solving some problem of interest on $\boldsymbol{X}$, (e.g. clustering). Classically, coresets have been proposed as a tool for solving optimization problems in settings where using the whole dataset is prohibitive or computationally intractable~\cite{agarwal2005geometric}. In classical machine learning, coreset techniques have been successfully used to reduce training times by creating coresets with gradients that closely approximate that of the full dataset ~\cite{mirzasoleiman2020coresets, sinha2020small}.

A variety of algorithms exist for constructing coresets~\cite{phillips2016coresets,campbellBayesianCoresetConstruction2018, campbellSparseVariationalInference2019}. Although recent work has proposed and explored using coresets for quantum algorithms where encoding the whole dataset is costly or infeasible~\cite{harrowSmallQuantumComputers2020, tomeshCoresetClusteringSmall2020}. This can be done \textit{statically}, where a classical computer generates the coreset which is then fed to the quantum algorithm and remains unchanged~\cite{tomeshCoresetClusteringSmall2020}. It's also been proposed to do coreset construction \textit{adaptively} where an iterative classical coreset algorithm queries a quantum computer to sample solutions to the problem on a coreset that is built up each iteration~\cite{harrowSmallQuantumComputers2020}.

\subsection{Boltzmann Machines}
One of the first machine learning models for the generative task of learning and sampling arbitrary probability distributions~\cite{hinton1986learning}, \textit{Boltzmann Machines} form the basis of other models such as deep belief networks~\cite{hinton2006fast}, and have been used for speech recognition~\cite{mohamed2011acoustic}, image generation~\cite{hinton2006fast}, and detecting network anomalies~\cite{fiore2013network}.

\begin{figure}
    \centering
    \includegraphics[width=0.85\linewidth]{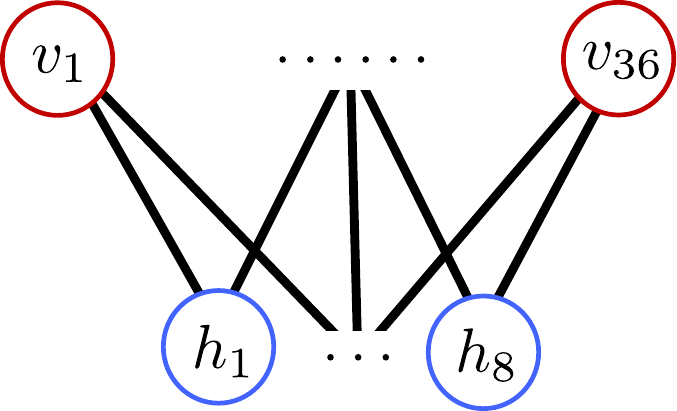}
    \caption{In our experiments we use QBMs with 36 visible units and 8 hidden units. Graphically, this is a fully connected bipartite graph where each node is a qubit/unit in our QBM. In terms of the Hamiltonian $H$, each graph node represents a $\sigma_a^z$ term with bias $b_a$ and graph edges $(a,b)$ represent $\sigma_a^z \sigma_b^z$ terms with weights $u_{ab}$.}
    \label{fig:qbm}
\end{figure}

A Boltzmann Machine is defined by a graph of binary units $z_a$ with biases $b_a$ connected by weighted edges $u_{ab}$. As shown in Figure~\ref{fig:qbm}, units can either be \textit{visible}, representing the input and output data, or \textit{hidden}, representing the model's internals.

A corresponding energy function to the graph is defined: 
\begin{equation}
    E(\boldsymbol{z}) = -\sum_{a}b_a z_a - \sum_{a,b}w_{ab}z_a z_b
\end{equation}
where $\boldsymbol{z}=(z_0,..., z_a,..)$ is a specific state of all the model's units, and $z_a \in \{+1, -1\}$. For convenience we also define $\boldsymbol{z} = (\boldsymbol{v},\boldsymbol{h})$ where $\boldsymbol{v}$ are the visible units and $\boldsymbol{h}$ are the hidden units.
Then, the Boltzmann Machine's learned distribution is the Boltzmann distribution for energy $E(\boldsymbol{z})$ summed over the possible states of the hidden units:
\begin{equation}
    P(\boldsymbol{v}) = \frac{1}{Z}\sum_{\boldsymbol{h}} e^{E(\boldsymbol{v}, \boldsymbol{h})}, Z=\sum_{\boldsymbol{v}}\sum_{\boldsymbol{h}} E(\boldsymbol{v}, \boldsymbol{h})
\end{equation}

The goal of training is to adjust parameters $b_a$ and $u_{ab}$ so that $P(\boldsymbol{v})$ approximates our training data set $P_{\text{data}}(\boldsymbol{v})$. This is equivalent to minimizing the negative log-likelihood
\begin{equation}
    \mathcal{L} = -\sum_{\boldsymbol{v}} P_{\text{data}}(\boldsymbol{v}) \log P(\boldsymbol{v})
\end{equation}
This can be done using a gradient based technique where the gradient with respect to the model parameters $\theta \in \{b_a, u_{ab}\}$ is
\begin{equation}
    \partial_{\theta} \mathcal{L} = \sum_{\boldsymbol{v}}P_{\text{data}}(\boldsymbol{v})\langle \partial_{\theta}E(\boldsymbol{v}, \boldsymbol{h})\rangle_{\boldsymbol{v}} - \langle \partial_{\theta}E(\boldsymbol{v}, \boldsymbol{h})\rangle
\end{equation}
where $\langle \dots \rangle_{\boldsymbol{v}}$, often called the \textit{positive phase}, is the expectation where the visible units are clamped to be the visible state $\boldsymbol{v}$, and $\langle \dots \rangle$, often called the \textit{negative phase}, is the unclamped expectation. 

To minimize $\mathcal{L}$ the model parameters can then be updated by taking a step in the direction of the negative gradient with some step size $\eta$
\begin{equation}
    \delta \theta = -\eta \partial_{\theta} \mathcal{L}
\end{equation}
Expressed for $b_a$ and $u_{ab}$ we have
\begin{equation}
    \delta b_a = -\eta \left(\overline{\langle z_a \rangle_{\boldsymbol{v}}} - \langle z_a \rangle\right),
\end{equation}
\begin{equation}
    \delta u_{ab} = -\eta \left(\overline{\langle z_a z_b \rangle_{\boldsymbol{v}}} - \langle z_a z_b \rangle\right).
\end{equation}
where $\overline{\langle \dots \rangle_{\boldsymbol{v}}} = \sum_{\boldsymbol{v}} P_{\text{data}}(\boldsymbol{v})\langle \dots \rangle_{\boldsymbol{v}}$ is the average expectation for a multiset of data $\{\boldsymbol{v_1}, \dots, \boldsymbol{v_n}\}$.

Calculating these gradient updates for general Boltzmann Machines can be exponentially costly and so approximate sampling-based methods are often employed. Additionally, a popular technique is to restrict the model graph to be a bipartite graph on the visible units and hidden units, called a \textit{Restricted Boltzmann Machine}, which allows efficient training through approximate methods like Contrastive Divergence~\cite{hinton2006fast}.

\subsection{Quantum Boltzmann Machines}
A \textit{Quantum Boltzmann Machine} (QBM) is a Boltzmann Machine where units are replaced by qubits and the energy function $E(\boldsymbol{z})$ is now a corresponding Transverse-field Ising model Hamiltonian~\cite{aminQuantumBoltzmannMachine2018}
\begin{equation}
    H = -\sum_a \Gamma_a \sigma^x_a -\sum_a b_a \sigma^z_a -\sum_{a,b} u_{ab} \sigma^z_a\sigma^z_b
\end{equation}
where $\sigma^i_a$ is the Pauli matrix $\sigma_i$ acting on qubit $a$. 

Defining $\Lambda_{\boldsymbol{v}}$ as the projector to the subspace with visible units equal to $\boldsymbol{v}$, our learned distribution is now:
\begin{equation}
    P(\boldsymbol{v}) = \Tr[\Lambda_{\boldsymbol{v}} \rho]
\end{equation}
where $\rho$ is the Gibbs state with partition function $Z$
\begin{equation}
    \rho = \frac{e^{-H}}{Z}, Z = \Tr[e^{-H}]
\end{equation}

Like the classical Boltzmann Machine case, our goal is to find model parameters $\boldsymbol{\theta} = (\boldsymbol{\Gamma}, \boldsymbol{b}, \boldsymbol{u})$ such that $P(\boldsymbol{v})$ approximates the data distribution $P_{\text{data}}(\boldsymbol{v})$.
Often an upper bound of the negative log-likelihood is optimized on to make training more tractable, however, doing so forces $\boldsymbol{\Gamma} \rightarrow \boldsymbol{0}$ and so $\Gamma_a = \Gamma$ becomes a hyperparameter fixed for all units. Similar to the classical case, gradient updates can then be calculated as
\begin{equation} \label{bias_grad}
    \delta b_a = -\eta \left(\overline{\langle \sigma_a^z \rangle_{\boldsymbol{v}}} - \langle \sigma_a^z \rangle\right),
\end{equation}
\begin{equation} \label{weight_grad}
    \delta u_{ab} = -\eta \left(\overline{\langle \sigma_a^z \sigma_b^z \rangle_{\boldsymbol{v}}} - \langle \sigma_a^z \sigma_b^z \rangle\right).
\end{equation}

Additionally, if our model is \textit{restricted} and has no connections between hidden units, then the positive phase for hidden units can be calculated exactly as
\begin{equation} \label{pos_phase}
    \langle \sigma_a^z \rangle_{\boldsymbol{v}} = \frac{b^{\text{eff}}_a(\boldsymbol{v})}{D_a(\boldsymbol{v})}\tanh{D_a(\boldsymbol{v})},
\end{equation}
\begin{equation}
    D_a(\boldsymbol{v}) = \sqrt{\Gamma_a^2 + (b^{\text{eff}}_a(\boldsymbol{v}))^2}
\end{equation}
where $b^{\text{eff}}_a(\boldsymbol{v}) = b_a + \sum_{b}u_{ab}\boldsymbol{v}_b$ is the effective bias on hidden unit $a$ based on the clamped state of each visible unit $b$ for visible data $\boldsymbol{v}$. For more background on training QBMs, we refer the reader to~\cite{aminQuantumBoltzmannMachine2018}.

Although we can calculate the positive phase cheaply, calculating the negative phase still requires Gibbs state sampling of our model Hamiltonian, which is expensive. Prior work has explored how to do this sampling variationally on near term quantum computers~\cite{zoufalVariationalQuantumBoltzmann2021} and proposed algorithms exist for Gibbs state sampling~\cite{chowdhury2016quantum, yung2012quantum, poulin2009sampling}, but finding the best way to perform Gibbs state sampling that is also tractable on current machines is still an active area of research.

\section{Motivation} \label{motivation}
Training a Boltzmann Machine requires a binary data set $\boldsymbol{X}=\{\boldsymbol{x_1},\dots,\boldsymbol{x_n}\}$ of size $n$. The task is to learn the probability distribution $P_{data}(\boldsymbol{v})$ corresponding to $\boldsymbol{X}$ where  $\boldsymbol{x}, \boldsymbol{v}$ are bitstrings of length $d$. It's typical to split $\boldsymbol{X}$ into small, constant-sized \textit{mini-batches} $\boldsymbol{B_i}=\{\boldsymbol{b_1},\dots,\boldsymbol{b_k}\}$ of size $k$ where $\boldsymbol{X} = \boldsymbol{B_1} \cup \dots \cup \boldsymbol{B_n}$~\cite{hintonPracticalGuideTraining2012}. 
The training procedure then entails iterating over $\boldsymbol{X}$ while performing a gradient-based update on our model parameters $\boldsymbol{\theta}$ for each mini-batch $\boldsymbol{B_i}$. One iteration through all our mini-batches is an \textit{epoch} and training can continue for enough epochs until $\boldsymbol{\theta}$ has sufficiently converged.

\begin{figure}
    \centering
    \includegraphics[width=\linewidth]{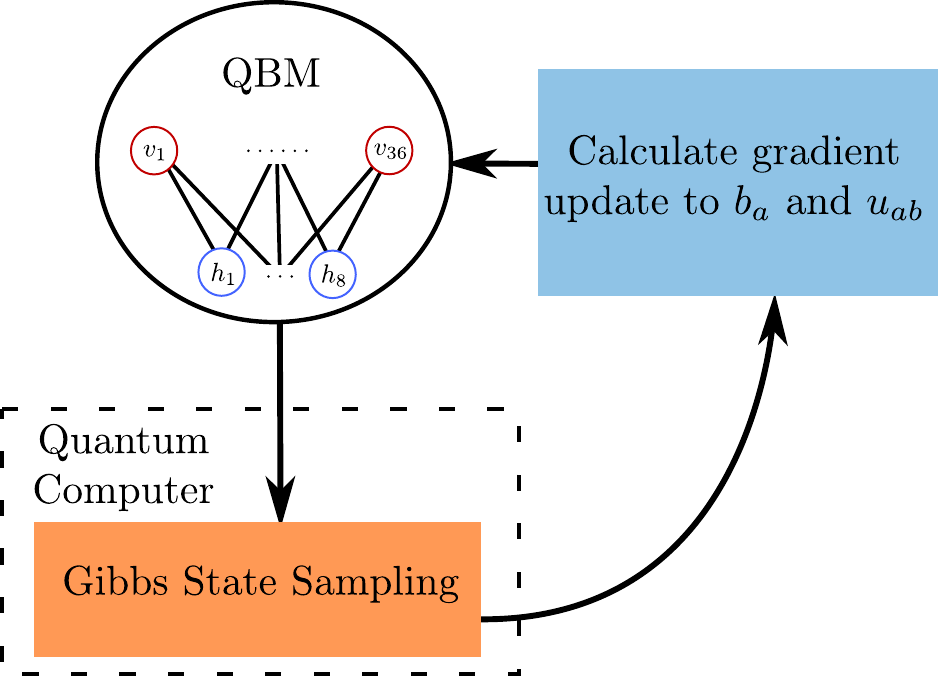}
    \caption{A high-level overview of the QBM training loop. We propose using a coreset in place of the full data set to minimize the number of iterations needed to reach a target model quality. In our experiments we're constrained to doing Gibbs state sampling using classical approximation techniques rather than using a quantum computer.}
    \label{fig:training}
\end{figure}

For a QBM, calculating the negative phase is the computational bottleneck since it requires Gibbs state sampling. For each mini-batch we need to calculate $|\boldsymbol{\theta}|$ negative phases. Conveniently since we only calculate $\sigma_z$ expectations, we can estimate all of these negative phases at once by just averaging the measured bitstrings, assuming we can approximately sample from the full Gibbs state. Therefore, we only need to do Gibbs state sampling once per mini-batch. For one epoch through the whole dataset $X$ this equates to $\frac{n}{k}$ instances of Gibbs state sampling. If we instead replace $X$ with a coreset $X'$ of size $m \ll n$ then we can substantially reduce the amount of times we need to perform Gibbs state sampling per iteration through the dataset. A high-level overview of this QBM training loop is shown in Figure~\ref{fig:training}. 

It's important to note that this also means we perform equally less parameter updates per epoch, and so it's necessary to choose $X'$ such that it sufficiently summarizes $X$. However, since constructing such $X'$ has been done successfully for classical machine learning models~\cite{mirzasoleiman2020coresets, sinha2020small}, we suspect it is also feasible for QBMs.

\section{Methodology}
\subsection{Bars $\times$ Stripes}
To apply coresets, we need a data set with: 1) a high enough dimensionality $d$ such that a perfect coreset of size $2^d$ cannot be constructed, and 2) $n$ data points where $n$ is large enough that we can create a coreset of size $m \ll n$.

\begin{figure}
    \centering
    \includegraphics[width=0.85\linewidth]{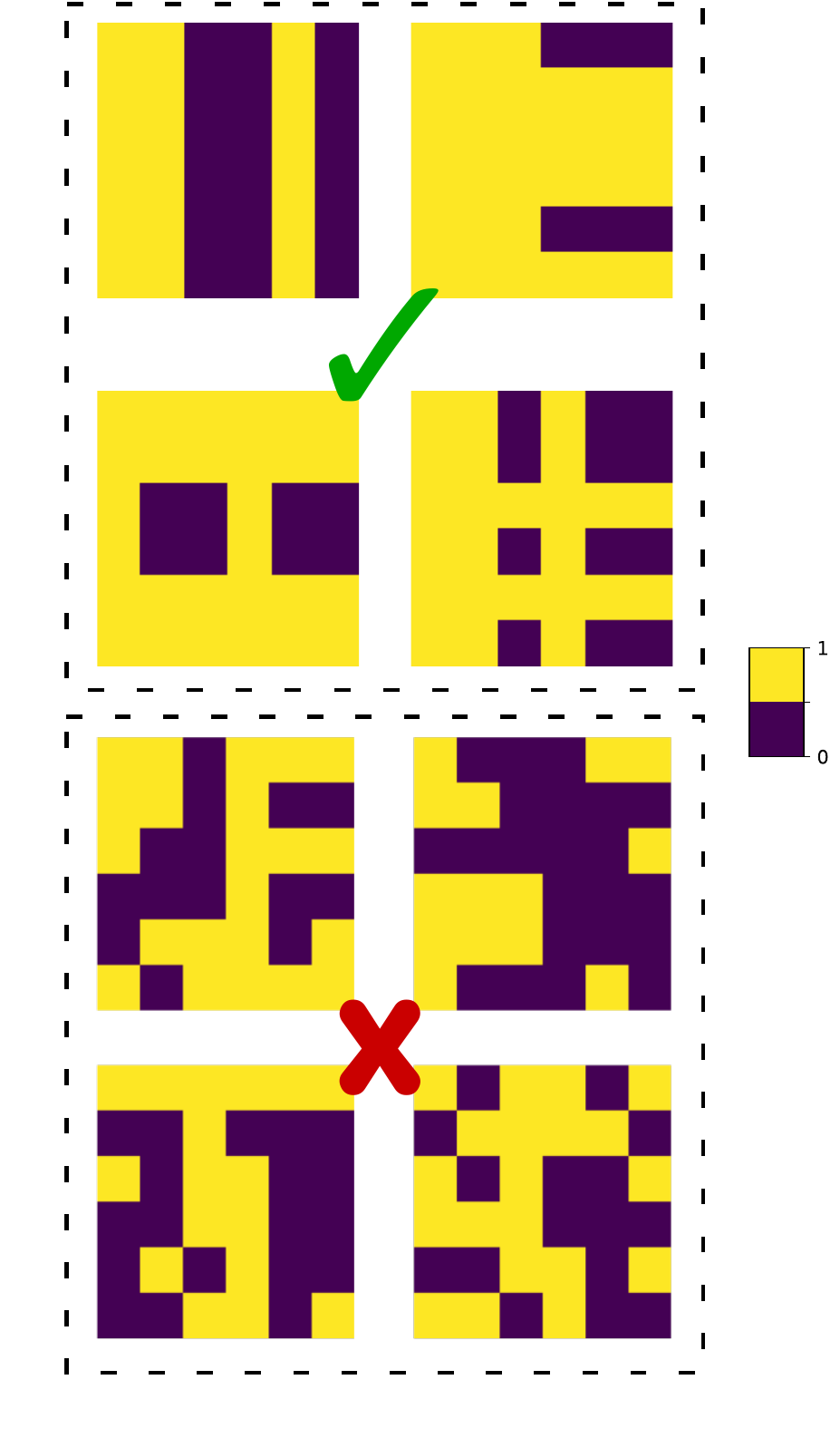}
    \caption{The Bars $\times$ Stripes (BXS) data set we use contains images made of only horizontal/vertical lines that span the full width/height of the 6x6 image. This equates to $\approx 4000$ distinct images in the data set.}
    \label{fig:bxs_dataset}
\end{figure}

In the past, work on smaller generative models, including QBMs~\cite{benedetti2019generative}, has used the Bars and Stripes (BAS) data set. In the normal definition of BAS, a data point is a binary $p \times q$ image consisting of only vertical lines \textbf{or} only horizontal lines.
To use this with coresets, we can choose $p,q$ such that $d = pq$ is sufficiently large to satisfy criteria 1, however, the number of distinct BAS images is only $O(2^p + 2^q)$, and so our data set might be too small to satisify criteria 2. One option is to just choose large enough $p,q$, however, we need $d$ visible units in our QBM and getting even $O(10^3)$ distinct images would require 81 visible units. This is too large to use with current methods, and so instead we opt to define images as only vertical lines \textbf{and} horizontal lines. To avoid confusion, we call this data set Bars $\times$ Stripes (BXS) as shown in Figure~\ref{fig:bxs_dataset}. With this formulation the number of distinct BXS images is $O(2^{p+q})$ and so we can construct a data set of $\approx4000$ distinct BXS images with only $p=q=6$, needing 36 visible units. Figure~\ref{fig:qbm} depicts the QBMs we use to learn this data set.

\subsection{Gibbs State Sampling}\label{gibbs}
As described in Section \ref{motivation}, training a QBM boils down to Gibbs state sampling, a task quantum computers are suspected to perform well at~\cite{poulin2009sampling}.
Unfortunately, given that we use QBMs with 44 units corresponding to 44 qubits, and efficiently performing Gibbs state sampling on current quantum computers is still an active area of research, it's difficult for us to use current quantum computers for Gibbs state sampling. Instead we opt to emulate previous work and use classical approximate methods~\cite{anschuetzNearTermQuantumClassicalAssociative2019}. 

To approximately sample from the Gibbs state we use population annealing with path integral Monte Carlo updates. Using the Trotter-Suzuki mapping, we approximate our quantum system with a corresponding classical system that has an additional imaginary time dimension discretized into $M$ slices~\cite{suzuki1976relationship}. As given in~\cite{anschuetzNearTermQuantumClassicalAssociative2019}, the corresponding probability distribution is then
\begin{equation}
    p_{\beta}(\boldsymbol{z^m}) \propto \exp{\left(-\beta [E_{\text{cl}}(\boldsymbol{z^m}) + E_{\text{qm}}(\boldsymbol{z^m};\beta)]\right)},
\end{equation}
where
\begin{equation}
    E_{\text{cl}}(\boldsymbol{z^m}) = \frac{1}{M}\sum_{m=1}^{M}E(\boldsymbol{z^m}),
\end{equation}
\begin{equation}
    E_{\text{qm}}(\boldsymbol{z^m};\beta) = \frac{1}{2\beta}\sum_{a,m}\ln\left(\tanh\left(\frac{\beta \Gamma_a}{M}\right)\right)z_a^m z_a^{m+1}
\end{equation}
In the above equations, $\boldsymbol{z^m}$ denotes the $m^\text{th}$ imaginary time slice of the system, and $z_a^m$ is the state of the $a^\text{th}$ binary unit in the $m^\text{th}$ imaginary time slice. Additionally, periodic boundary conditions are enforced so that $m=M+1$ is $m=1$.

In population annealing, we maintain $K$ replicas of an initial state of this classical system and iterate through increasing values of $\beta = \frac{1}{T}$, the inverese temperature. Each iteration the population of replicas is resampled based on their relative Boltzmann weights and replicas are updated by a finite number of Monte Carlo steps~\cite{hukushimaPopulationAnnealingIts2003}. Afterwards, we can treat the first imaginary time slice of each replica as an approximate sample from our Gibbs state.

\subsection{QBM Training}

In our numerical experiments we take $\Gamma_a = 2$ and optimize the biases and weights using the Adam method~\cite{kingma2014adam} with the gradient calculations from Eq.~\ref{bias_grad} and Eq.~\ref{weight_grad}. We split the input data set into mini-batches $\boldsymbol{B_i} = \{\boldsymbol{b_1}, \dots, \boldsymbol{b_k}\}$ of size $k=32$. Negative phases are calculated using the procedure described in Section~\ref{gibbs} with 128 replicas, 5 iterations from $\beta = 0$ to $\beta = 1$, and $M=10$ imaginary time slices. For the positive phases, since a coreset has weighted points the exact calculation becomes
\begin{equation}
    \overline{\langle \sigma_a^z\rangle_{\boldsymbol{b}}} = \frac{1}{\sum_j w_j}\sum_j w_j \langle \sigma_a^z\rangle_{\boldsymbol{b_j}}
\end{equation}
using Eq.~\ref{pos_phase} for $\langle \sigma_a^z\rangle_{\boldsymbol{b_j}}$. 

We initialize our weights and biases using the recipe described in~\cite{hintonPracticalGuideTraining2012}. Weights are sampled from a normal distribution centered at 0 with a standard deviation of 0.01, and biases are calculated to be $b_a = \log[p_a / (1 - p_a)]$, where $p_a$ is the proportion of training data where bit $a$ is on.

\subsection{QBM Evaluation}\label{evaluation}
Since our goal is for $P(\boldsymbol{v})$ to approximate $P_{\text{data}}(\boldsymbol{v})$ we might want to evaluate the KL divergence between the two distributions to see how accurate the model is
\begin{equation}
    KL = \sum_{\boldsymbol{v}} P_{\text{data}}(\boldsymbol{v})\log{\frac{P_{\text{data}}(\boldsymbol{v})}{P(\boldsymbol{v})}}
\end{equation}
However, for high dimensionality data this is typically impractical to calculate, since we only know empirical distributions for $P_{\text{data}}(\boldsymbol{v})$ and $P(\boldsymbol{v})$ from the data set and our Gibbs state sampling, respectively. Since the possible values of $\boldsymbol{v}$ scales exponentially with the dimension it's unlikely these empirical distributions are large enough to have meaningful overlap where we can calculate the KL divergence. Additionally, in our experiments where we know the underlying data distribution is the $\approx 4000$ BXS images, we still only get 128 samples to estimate our model distribution, and so the KL divergence will always be $\infty$.

To avoid this issue and allow for a distinction between "close" images (e.g. an image that is 1 pixel from a BXS image) and other incorrect images, we adopt a strategy similar to the inception score~\cite{salimans2016improved}. We train a classical feedforward neural network with 3 layers to differentiate BXS images from non-BXS images to $>99\%$ validation accuracy. We then have this predict our QBM samples and use the BXS classification score as a proxy for the quality of our QBM samples. 

It's important to note that in our experiments we don't have multiple classifications for BXS images, and so we can't replicate the actual inception score which intuitively also checks that generated images come from a diverse number of classes to discount model overfitting. As a result, our metric is susceptible to being fooled by a model which has overfit to only a couple of images. However, we still find it to be a useful metric to evaluate QBM training.

\subsection{Coreset Construction}
In our preliminary experiments we create two types of coresets of size $m=128$. The first is simply a uniform sampling of $m$ images from the data distribution. The second is constructed by solving the minimax facility location problem
\begin{equation}
    \boldsymbol{X'} = \arg \min_{\boldsymbol{X'}} \left( \max_{\boldsymbol{x_i} \in \boldsymbol{X}} \min_{\boldsymbol{x'_j} \in \boldsymbol{X'}}d(\boldsymbol{x_i},\boldsymbol{x'_j})\right)
\end{equation}
similarly to~\cite{sinha2020small}. Intuitively, solving this problem entails choosing a coreset $X'$ such that we minimize the maximum distance for a point in the full dataset $X$ to its closest point in the coreset. In this formulation $d(\boldsymbol{x_i}, \boldsymbol{x'_j})$ is some distance function between data set point $\boldsymbol{x_i}$ and coreset point $\boldsymbol{x'_j}$. Solving this problem is NP-Hard~\cite{fowler1981optimal} and so we solve it greedily using Algorithm 1 from~\cite{sinha2020small}. 

Like~\cite{sinha2020small}, we don't use the Euclidean distance between two data points in the $d=pq$ dimensional space for $d(\boldsymbol{x_i}, \boldsymbol{x'_j})$, which is often not meaningful for images or useful in high-dimensional spaces. Instead, we reduce each data point to 8 dimensions using the output of the second to last layer of the classical neural network from Section~\ref{evaluation} and calculate Euclidean distances in this 8 dimensional space. We refer to this as the Inception Distance (ID), with the intuition that this projection will exploit more semantic information and be of a small enough dimensionality to get useful values for $d(\boldsymbol{x_i}, \boldsymbol{x'_j})$.

\section{Results}
\begin{figure}
    \centering
    \includegraphics[width=\linewidth]{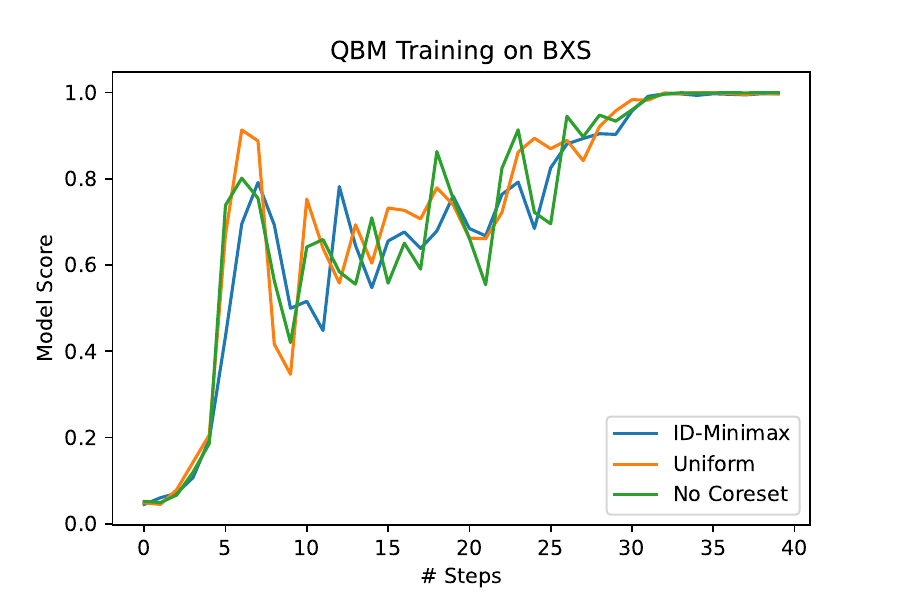}
    \caption{Results of our preliminary experiments training QBMs with and without coresets using the BXS dataset described in Section~\ref{fig:bxs_dataset}. The model score corresponds to the average BXS classification score from Section~\ref{evaluation} over the 128 samples from Gibbs state sampling of the model distribution $P(\boldsymbol{v})$. A higher score means the samples are on average closer to the BXS data set, reflecting the QBM's learning progress. In this figure, each line is the averaged scores over 10 experiments.}
    \label{fig:results}
\end{figure}

We ran initial numerical experiments comparing uniform coresets, inception distance minimax coresets, and no coresets for a QBM with 36 visible units and 8 hidden units learning the BXS data set. The coresets were of size $m = 128$ and the full data set is of size $n=4096$. We designed our experiments with the idea that the user has some budget number of times to perform Gibbs state sampling, and so we run all experiments for 40 gradient-based updates, equalling 40 Gibbs state samplings. We use mini-batches of size $k = 32$, meaning with 40 iterations we go through 10 epochs with coresets and less than half an epoch with the full data set. Training was scored at each update step by averaging the BXS classification score for each of the 128 samples obtained from Gibbs state sampling. Figure~\ref{fig:results} shows the results of the experiments averaged over 10 runs for each approach. In the plots we can see that all three approaches are able to learn the data set successfully, validating our training methodology. However, using coresets did not improve the training time. We discuss hypotheses for why this occurs and resulting ideas for future work in Section~\ref{discussion}.

\section{Discussion} \label{discussion}
In this work we proposed using coreset techniques to reduce training times of Quantum Boltzmann Machines (QBMs). Coresets have already been used to reduce training times of classical machine learning models~\cite{mirzasoleiman2020coresets, sinha2020small}. In the case of QBMs, however, training is bottlenecked by Gibbs state sampling and so training time is equivalent to how often this sampling is performed. Since Gibbs state sampling is believed to be a promising use case for quantum computers~\cite{poulin2009sampling}, reducing the amount of times this sampling is needed equates to reducing the runs of the quantum hardware. In a scenario where computational time on a quantum computer is a precious resource, we propose that this might lead to substantial practical savings. Additionally, in a regime of noisy quantum computers with imperfect Gibbs state sampling algorithms, this reduction might also lead to less noisy results and better trained QBMs.

We performed initial numerical experiments exploring this direction. Although we expected to see similar results as work that used coresets for classical machine learning models~\cite{mirzasoleiman2020coresets, sinha2020small}, we find that, as shown in Figure~\ref{fig:results}, all approaches learn at the same rate. One possible explanation is that although we tried to maximize the problem size we could work with given our resources, the problem dimensionality and data set size are still too small, and the QBM trivially fits to the first couple mini-batches regardless of data set size. Another possibility is the BXS data set isn't diverse enough to have substantially differing positive phases for batches of size $k = 32$, meaning all mini-batches from the data set elicit roughly the same gradient update of the QBM parameters. Additionally we only look at two unweighted coreset constructions. It's possible they aren't able to coherently summarize the full dataset and enable more effective gradient updates. 

Despite these results, we still think coresets present a promising direction for practically training QBMs, and leave further experiments using weighted coreset constructions with larger models and data sets to future work.

\section*{Acknowledgment}
This work is funded in part by EPiQC, an NSF Expedition
in Computing, under award CCF-1730449; in part
by STAQ under award NSF Phy-1818914; in part by NSF
award 2110860; in part by the US Department of Energy Office 
of Advanced Scientific Computing Research, Accelerated 
Research for Quantum Computing Program; and in part by the 
NSF Quantum Leap Challenge Institute for Hybrid Quantum Architectures and 
Networks (NSF Award 2016136) and in part based upon work supported by the 
U.S. Department of Energy, Office of Science, National Quantum 
Information Science Research Centers. E.R.A. is supported by the National Science Foundation Graduate Research Fellowship Program under Grant No. 4000063445, and a Lester Wolfe Fellowship and the Henry W. Kendall Fellowship Fund from M.I.T.

FTC is Chief Scientist for Quantum Software at ColdQuanta and an advisor to Quantum Circuits, Inc.

\printbibliography

\end{document}